\documentclass{Interspeech2024}

\usepackage{multirow,siunitx}
\newcommand{\scell}[2][c]{%
  \begin{tabular}[#1]{@{}c@{}}#2\end{tabular}}
\interspeechcameraready

\title{Dual-Pipeline with Low-Rank Adaptation for\\New Language Integration in Multilingual ASR}

\name{Yerbolat Khassanov, Zhipeng Chen, Tianfeng Chen, Tze Yuang Chong, Wei Li,\\Jun Zhang, Lu Lu, Yuxuan Wang}

\address{ByteDance}
\email{\{yerb.khass, chenzhipeng.speech, tianfeng.chen, tychong, liwei.speech, zhangjun.jarry, lulu0314, wangyuxuan.11\}@bytedance.com}

\begin{document}

\maketitle

\begin{abstract}
This paper addresses challenges in integrating new languages into a pre-trained multilingual automatic speech recognition (mASR) system, particularly in scenarios where training data for existing languages is limited or unavailable. The proposed method employs a dual-pipeline with low-rank adaptation (LoRA). It maintains two data flow pipelines—one for existing languages and another for new languages. The primary pipeline follows the standard flow through the pre-trained parameters of mASR, while the secondary pipeline additionally utilizes language-specific parameters represented by LoRA and a separate output decoder module. Importantly, the proposed approach minimizes the performance degradation of existing languages and enables a language-agnostic operation mode, facilitated by a decoder selection strategy. We validate the effectiveness of the proposed method by extending the pre-trained Whisper model to 19 new languages from the FLEURS dataset.

%Experimental validation on the Whisper model, extended to 19 new languages from the FLEURS dataset, demonstrates a substantial X\% relative character error rate improvement over zero-shot performance, and outperforms two strong baseline systems by X\% and X\%, respectively.
\end{abstract}

\section{Introduction}

Recently, large-scale and massively multilingual automatic speech recognition (mASR) models have gained prominence in the speech community~\cite{DBLP:conf/icml/RadfordKXBMS23,zhang2023google,pratap2023scaling}. 
Typically pre-trained on extensive amounts of unsupervised data, followed by fine-tuning using supervised and/or weakly-supervised data from both publicly available and proprietary sources, these models demonstrate exceptional robustness to diverse audio conditions and exhibit broad generalization across domains, tasks, and languages.
Furthermore, they achieve state-of-the-art performance in the majority of benchmarks\footnote{\url{https://huggingface.co/spaces/hf-audio/open_asr_leaderboard}}, leading to high popularity among both academia and industry practitioners.
Importantly, most of these mASR models are publicly shared\footnote{\url{https://huggingface.co/models}} with the intention to make speech technology more accessible and to further advance innovation in speech-based applications.

%What is language extension in mASR, and why do we need it? Why it is challenging?
Nonetheless, extending large mASR models to new languages demands significant computational resources, involving multiple iterations of re-training with adjusted hyperparameters and potentially modified architecture.
Furthermore, in certain application scenarios, access to training data for existing languages may be restricted or entirely absent, particularly with publicly shared models that are often distributed without accompanying training data.
Consequently, refining these models while preserving comparable performance on existing languages presents a substantial challenge. 
This challenge is further heightened under a language-agnostic scenario, where the language ID of the input utterance is unknown—an often encountered situation in real-world applications.

%How about other existing solutions? Why they do not work?
In this context, parameter-efficient fine-tuning techniques~\cite{DBLP:conf/iclr/HeZMBN22}, such as adapters~\cite{DBLP:conf/icml/HoulsbyGJMLGAG19,DBLP:conf/acl/MahabadiR0H20}, are less effective as they can result in catastrophic forgetting of existing languages.
Traditional language integration techniques, like continual learning~\cite{DBLP:journals/nn/ParisiKPKW19,DBLP:conf/icassp/LiPZSSHZFGP22}, become impractical due to the need for training data from existing languages.
On the other hand, straightforward solutions, such as maintaining a separate copy of the large mASR model for each group of languages (potentially preceded by a language identification model), not only incur higher computational and storage resource requirements but also forfeit other benefits offered by multilingual models~\cite{li2021scaling,pratap20c_interspeech}.

\begin{figure}[t!]
    \centering
    \includegraphics[width=0.65\linewidth, trim={0cm 4.0cm 0cm 4.0cm}, clip=true]{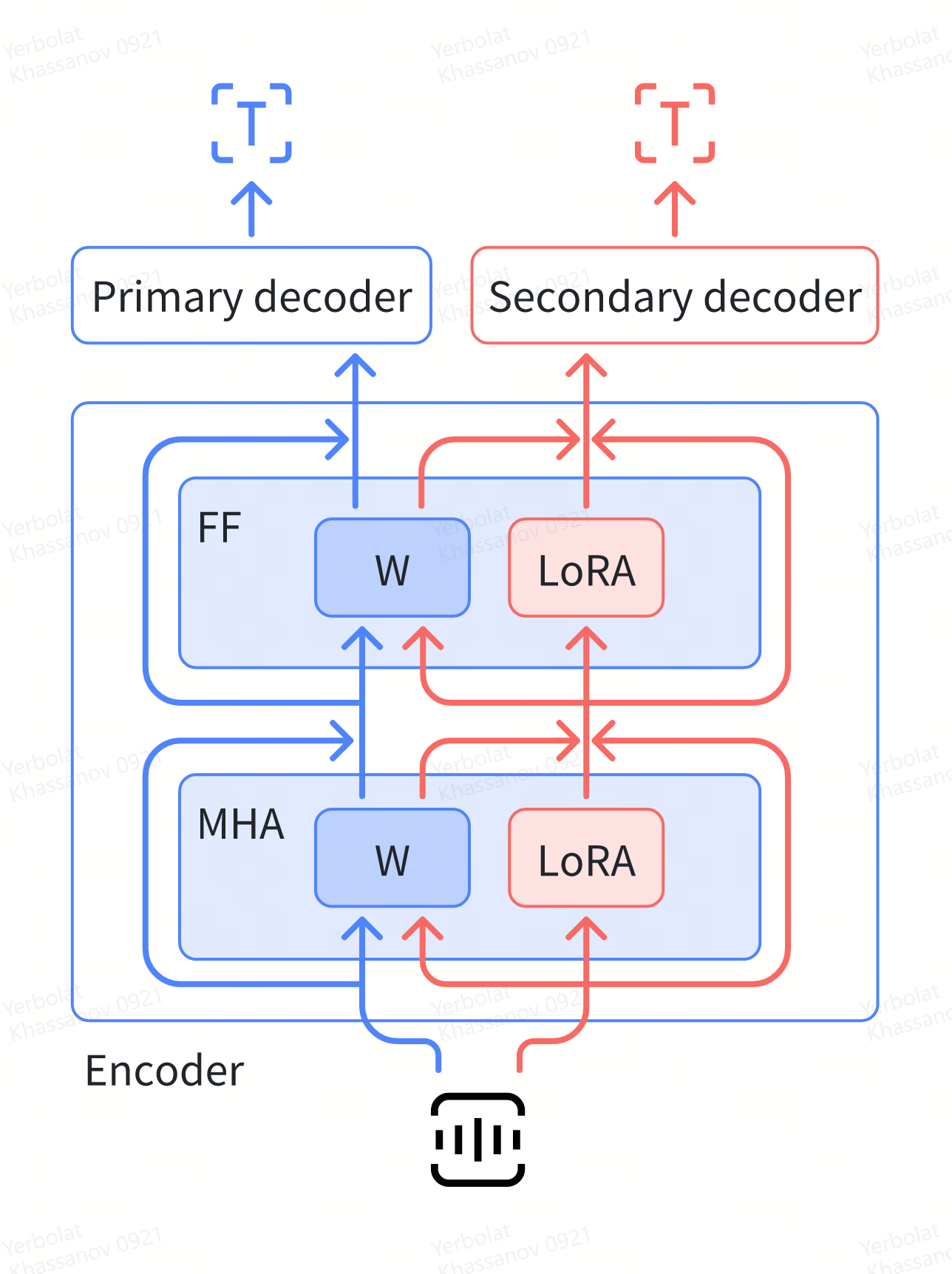}
    %\vspace{-0.2cm}
    \caption{Architecture of Transformer-based multilingual ASR employing dual-pipeline with LoRA.}
    \label{fig:arch}
    %\vspace{-0.5cm}
\end{figure}

%What do you propose and how it addresses the aforementioned drawback?
In this study, we propose a dual-pipeline with low-rank adaptation (LoRA)~\cite{DBLP:conf/iclr/HuSWALWWC22} approach to effectively integrate new languages into an mASR (Figure~\ref{fig:arch}). 
Unlike other language extension methods, our approach does not depend on the training data of existing languages.
To achieve this, our method maintains two separate data flow pipelines. 
The primary pipeline is for existing languages, and it passes through the pre-trained parameters of mASR, which are always kept unchanged. 
The secondary pipeline is dedicated to new languages and, in addition to pre-trained parameters, it leverages language-specific parameters represented by LoRA and a separate decoder module. 
Overall, the proposed method introduces a minimal number of additional parameters and is computationally efficient.

To evaluate the effectiveness of our method, we extend the publicly available Whisper (Large-V2) model~\cite{DBLP:conf/icml/RadfordKXBMS23} to 19 previously unseen languages from the FLEURS dataset~\cite{DBLP:conf/slt/ConneauMKZADRRB22}.
Experimental results demonstrate that our method achieves significant improvements over the zero-shot and other two strong baselines.
Additionally, we demonstrate its performance in a language-agnostic mode using a simple decoder selection strategy across 102 languages of the FLEURS dataset.

\iffalse
The remainder of the paper is organized as follows. 
Section two provides a brief overview of related work.
Section three details the experimental setup, presents obtained results and conducted analyses. 
Finally, section four concludes the paper.
\fi

\section{Integrating new languages}
\subsection{Secondary decoder for new languages}
To expand the mASR model to incorporate new languages, we introduce a secondary decoder component that facilitates the output token units for new languages (depicted in Figure~\ref{fig:arch}).
This methodology is akin to the multi-head architecture explored in prior research, where a shared encoder serves multiple decoders~\cite{DBLP:conf/icassp/LiPZSSHZFGP22,li2021scaling}. 
However, in contrast to our objective, those studies focus on fostering cross-lingual collaboration among linguistically similar languages by allocating each language group to a separate decoder.

The secondary decoder is initialized randomly and can be modeled using any network architecture. 
In our study, we chose an LSTM network to enhance decoding speed~\cite{zhang2018accelerating}.
The LSTM-based decoder is utilized alongside a 2-head additive attention mechanism, forming a Listen, Attend, and Spell (LAS) framework~\cite{DBLP:conf/icassp/ChanJLV16}. 
It's important to note that for each pipeline, we apply distinct final layer normalization~\cite{ba2016layer} before passing the encoder outputs to the decoders. 
The output format of the secondary decoder mirrors the structure of the primary decoder, where we first predict the unique language tag followed by the transcript.

\subsection{Dual-pipeline with LoRA}
The secondary decoder for new languages acts as a language model (LM), conditioned on both the previous context and the output features of the encoder component.
Since the parameters of the encoder component remain unchanged to preserve performance in existing languages, it lacks exposure to new languages, potentially leading to suboptimal performance.
To address this issue, we propose maintaining two separate pipelines for existing and new languages.
However, dedicating a distinct pipeline with its own parameters is akin to having a new encoder, introducing computational inefficiencies.

Therefore, we employ the recently proposed LoRA technique~\cite{DBLP:conf/iclr/HuSWALWWC22}, which introduces trainable low-rank matrices $A$ and $B$ to efficiently adapt model parameters to new domain. 
Specifically, we apply LoRA to all pre-trained weight matrices $W$ in multi-head attention (MHA) and feed-forward (FF) sub-layers of the Transformer-based encoder, as depicted in Figure 1. 
This results in the following computation:
\begin{align}
  h &= Wx+BAx
\end{align}
As a result, the secondary pipeline leverages important feature transformations from pre-trained matrices, along with the language-specific LoRA module. 
While it is possible to allocate a separate LoRA module for each new language, in this work, we choose to utilize a single LoRA module for all new languages. 
Additionally, we maintain separate residual connections~\cite{DBLP:conf/cvpr/HeZRS16} for the secondary pipeline.

During the fine-tuning stage, we exclusively update the parameters of the secondary decoder and LoRA using data from new languages. 
In the decoding stage, unlike the original LoRA technique, we avoid merging the LoRA with the pre-trained weight matrices.

\subsection{Decoder selection strategy}
\label{sec:decoder_selection}
The challenge inherent in employing multiple decoders lies in determining the final recognition output without prior knowledge of the input audio's language ID.
To tackle this issue, we employ a decoder selection strategy~\cite{khassanov_icassp24}, facilitating a fully language-agnostic mode. 
In this strategy, we first compare log-probability scores of identified language tags from each decoder for a given input audio. 
If the difference falls below a predetermined threshold ($\tau$), we proceed to compare the average log-probability scores of the full transcripts. 
Adjusting this threshold allows us to manage decoding speed. 
A smaller threshold, for instance, enables decision-making without calculating scores for the remaining tokens using both decoders.
Moreover, we introduced a bias score ($\beta$) added to the average log-probability score of the secondary decoder, enabling the prioritization of one decoder over the other. 
Although the current strategy could be further refined, it is sufficient for our present demonstration.

\section{Experiment}

\subsection{Experimental setup}
\textbf{mASR.}
We utilized the Whisper (Large-V2)~\cite{DBLP:conf/icml/RadfordKXBMS23} model, a multi-task and multilingual speech processing system with 1.5 billion parameters, employing an encoder-decoder Transformer~\cite{DBLP:conf/nips/VaswaniSPUJGKP17} network.
The model is trained on a diverse dataset of 680,000 hours, encompassing various speech processing tasks like multilingual speech recognition, speech translation, spoken language identification, and voice activity detection.
This dataset comprises audio paired with transcripts sourced from the Internet, ensuring a wide distribution across different environments, recording setups, speakers, and languages.
Throughout the experiments, we maintain the parameters of the Whisper model unchanged to preserve its performance in speech recognition for existing languages and other tasks.

\textbf{Data.}
We conducted experiments on 19 languages selected from the FLEURS\footnote{Cantonese was not included in our experiments because, in the Whisper model, all varieties of Chinese were amalgamated under a single language code \texttt{<|zh|>}, potentially encompassing other dialects.} dataset~\cite{DBLP:conf/slt/ConneauMKZADRRB22}, as specified in Table~\ref{tab:cer_results}. 
Each language is represented by approximately 10 hours of training data and has not been encountered by the Whisper model previously. 
The output vocabulary for these languages is formed from the unified text using a byte-level byte pair encoding (BPE) algorithm~\cite{DBLP:conf/aaai/WangCG20}, with a size set to 2,000.

\textbf{Decoder and LoRA components for new languages.}
The secondary decoder was implemented using a single-layer LSTM with 512 hidden units, unless specified otherwise. 
In the case of LoRA, we explored various rank values, and tuning the corresponding scaling factor $\alpha$~\cite{DBLP:conf/iclr/HuSWALWWC22} was necessary. 
We found that $\alpha$ values from \{$1, 2, 4, 8$\} worked best in our experiments.

\textbf{Fine-tuning.}
We aggregate data from all 19 languages and fine-tune the secondary decoder and LoRA components for 20k steps using 16 V100 GPUs. For models with over 100M trainable parameters, we used 4 A100 GPUs, maintaining the effective batch size unchanged.
The Adam optimizer~\cite{DBLP:journals/corr/KingmaB14} is employed, and we explore various learning rates from 
\{$1\times10^{-4},3\times10^{-4},5\times10^{-4},7\times10^{-4}$\}.
Our implementation incorporates a tri-stage learning rate schedule, including a warm-up for the initial 10\% of steps, a constant rate for the subsequent 40\% of steps, and decay during the final 50\% of steps. 
We select the last checkpoint as the final model.

\textbf{Evaluation.}
In all experiments, we used character error rate (CER) as an evaluation metric, and set the beam size to five. 
For all test sets, we applied our in-house voice activity detection (VAD) model to segment the utterances into audio chunks not exceeding 30 seconds. 
While this slightly hurt the CER performance, it helped to avoid long-form decoding heuristics~\cite{DBLP:conf/icml/RadfordKXBMS23}.
Similar to previous works~\cite{DBLP:conf/icml/RadfordKXBMS23,pratap2023scaling}, we applied the Whisper normalization\footnote{English normalization is applied to the English language, while basic normalization is used for other languages.} on reference and recognized output text before CER computation.
In the input prompt, we provided the \texttt{<|transcribe|>} and \texttt{<|notimestamps|>} special tokens, but did not include the ground-truth language tag token, assuming the language-agnostic scenario.
Consequently, we did not use the previous-text conditioning method to switch between traditional or simplified Chinese writing systems.
Additionally, we report the number of additional parameters introduced by appending the secondary decoder and LoRA.

\subsection{Experiment results}
We first evaluate the effectiveness of the proposed dual-pipeline with LoRA method in integrating new languages.
Subsequently, we adopt the group-aware scenario, assuming prior knowledge of the input audio's group (new or existing) and deploying the corresponding decoder.
Consequently, we present the average CER results for 19 new languages.
Note that in group-aware scenario, the performance of existing languages do not change.
The dual-pipeline starts from the input to the initial layer of the encoder network, applying LoRA to all parameters of the encoder, including the MHA and FF sub-layers.
We explore the implications of using different rank values for LoRA.

\begin{figure}[t]
    \centering
    %\vspace{-0.2cm}
    \includegraphics[width=0.95\linewidth, trim={0.6cm 1.9cm 0.4cm 0.7cm}, clip=true]{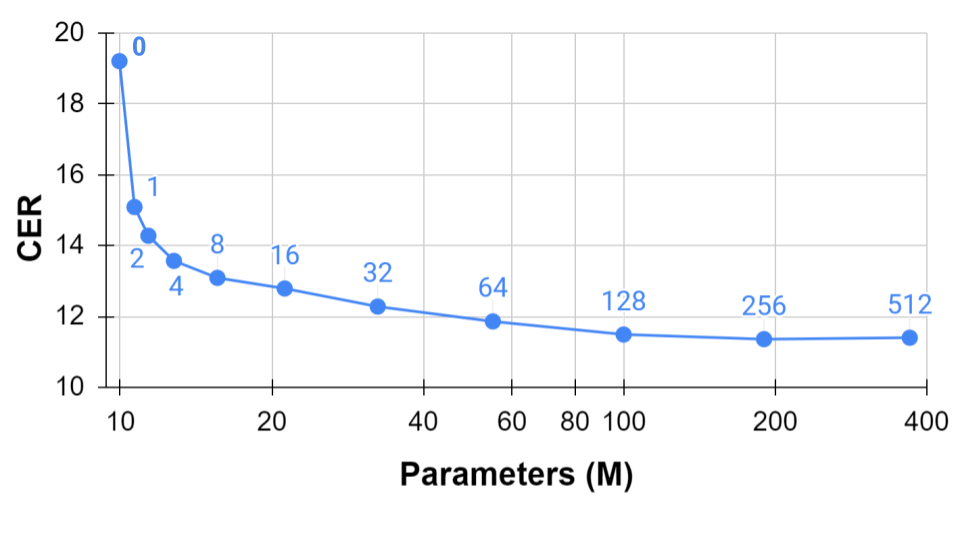}
    %\vspace{-0.3cm}
    \caption{Number of additional parameters and average CER results for 19 new languages integrated using our dual-pipeline with LoRA method. Data labels indicate the rank values used in LoRA component.}
    \label{fig:cer_rank}
    \vspace{-0.3cm}
\end{figure}

The experimental results in Figure~\ref{fig:cer_rank} indicate that increasing the rank generally improves CER performance at the cost of increasing the additional parameter size.
The best average CER of 11.36\% is achieved at rank 256.
However, we found the CER performance to converge starting from rank 128.
Further increasing the rank value significantly increases the additional parameter size without bringing any substantial CER improvement. %We also report the zero-shot performance of the pre-trained Whisper model, where no additional language-specific components or training steps were added.
%Additionally, we present the decoder-only results, omitting the LoRA component by setting the rank value to 0.
%The zero-shot performance of the pre-trained Whisper model on 19 new languages achieves average CER of 81.62\%.
The decoder-only setup, where we omitt the LoRA component by setting the rank value to 0, achieves 19.21\% average CER with 9.96M additional parameters.
%We observed that our method even with the rank of 1 significantly outperforms both zero-shot and decoder-only baselines, achieving 15.09\% average CER with 10.7M additional parameters.
We observed that our method even with the rank of 1 significantly outperforms the decoder-only baseline, achieving 15.09\% average CER with 10.7M additional parameters.
These results demonstrate the effectiveness of the proposed method.

\subsubsection{Starting from the intermediate layers}
In contrast to previous experiments, where the dual-pipeline initiates from the first layer, in these experiments, we commence it from the intermediate layers of the encoder network. 
This approach is motivated by the observation that the bottom layers of the encoder are more language-invariant and, therefore, can be shared across different languages. 
Moreover, this strategy reduces the number of additional parameters.
Whisper's encoder comprises 32 transformer layers, and we will investigate the impact of starting the dual-pipeline from layers \{$8, 16, 24, 28, 30$\}.
For these experiments, we will utilize LoRA with ranks set to 128, 256 and 512.
All experiments are conducted within the context of the group-aware scenario.

%The experimental results in Table 1 reveal that initiating the dual-pipeline from intermediate layers remains effective.
The experimental results in Table 1 reveal that the dual-pipeline initiated from the intermediate layers remains effective.
Moreover, initiating it from layers 8-16 does not compromise the CER performance while substantially reducing the number of additional parameters.
For instance, with a rank of 512, starting from the 16th layer improved the average CER by approximately 2\% relative (from 11.40\% to 11.18\%) and reduced the number of parameters for LoRA by 50\%.

\begin{table}[t]
    \caption{Average CER results for 19 new languages for different intermediate layers of encoder used to start the dual-pipeline with LoRA.}
    \label{tab:cer_inter}
    \vspace{-0.2cm}
    \centering
    \small
    \renewcommand\arraystretch{1.0}
    \setlength{\tabcolsep}{2.0mm}
    \begin{tabular}{l|cccccc}
        \toprule
        \multirow{2}{*}{rank}   & \multicolumn{6}{c}{Dual-pipeline starting layer} \\\cline{2-7}
                                & 0     & 8     & 16    & 24    & 28    & 30 \\
        \midrule
        128                     & 11.49 & \textbf{11.46} & 11.67 & 12.97 & 14.32 & 15.24 \\
        256                     & \textbf{11.36} & 11.38 & 11.47 & 12.55 & 13.96 & 15.11 \\
        512                     & 11.40 & 11.25 & \textbf{11.18} & 12.34 & 13.55 & 15.09 \\
        \bottomrule
    \end{tabular}
    \vspace{-0.5cm}
\end{table}

\subsubsection{Comparison to other baselines}
This section provides a comparative analysis of the proposed method with two strong baseline methods. 
Similar to previous experiments, we continue to utilize the group-aware scenario.
All methods were constrained to incorporate fewer than 30 million additional parameters, accounting for less than 0.1\% additional parameters per language.
The initial baseline employs a decoder-only approach, as described previously.
However, in these experiments, a greater number of parameters are allocated to enhance its performance.
We examined various combinations of layers and hidden units, concluding that a four-layer LSTM decoder with 512 hidden units yielded optimal results.

The second baseline comprises a supplementary encoder and decoder architecture~\cite{khassanov_icassp24}.
In this configuration, alongside a distinct decoder, an extra encoder component is incorporated for new languages.
This new encoder component is initialized from the original encoder and attached in a manner that maintains the total encoder depth at the same level, i.e., 32 layers.
Due to the parameter constraint, only a single-layer LSTM decoder with 512 units and a single-layer encoder could be added to the output of the 31st layer of the original encoder.

Lastly, our approach employs a dual-pipeline with LoRA initiated from the first layer.
The LoRA rank is set to 32.
The secondary decoder consists of a single-layer LSTM with 512 units.
Additionally, we report the zero-shot performance, enabled by the use of byte-level BPE tokens as output units in Whisper model.

%Lastly, our method uses dual-pipeline with LoRA starting from the first layer and rank set to 32.
%We also include the zero-shot performance results of the Whisper model, which doe not introduce any additional parameters, for demonstrative purposes.

The experiment results in Table~\ref{tab:cer_baseline} demonstrate that all baselines exhibit significant improvement over the zero-shot performance, with an average CER improvement of over 80\% relative. Among them, our method attains the best average CER of 12.79\%, marking a 23\% and 14\% relative improvement over the decoder-only and supplementary encoder-decoder baselines, respectively. Notably, this enhancement is achieved with the least number of additional parameters. 
The results demonstrate the superiority of our approach over other strong baselines within the constraints of the additional parameter size.

\begin{table}[t]
    \caption{Number of additional parameters and average CER results for 19 new languages obtained from zero-shot, decoder-only, supplementary encoder and decoder, and our dual-pipeline with LoRA approaches.}
    \label{tab:cer_baseline}
    \vspace{-0.2cm}
    \centering
    \small
    \renewcommand\arraystretch{0.9}
    \setlength{\tabcolsep}{2.7mm}
    \begin{tabular}{lcc}
        \toprule
                            & Additional param. & Average CER \\
        \midrule
        Zero-shot           & 0                 & 81.62 \\
        Decoder-only        & 23.47M            & 16.69 \\
        Sup. enc. \& dec.   & 28.73M            & 14.88 \\
        Ours                & 21.21M            & \textbf{12.79}\\    
        \bottomrule
    \end{tabular}
    %\vspace{-0.1cm}
\end{table}

\subsubsection{Language-agnostic mode with decoder selection}

Finally, we present the overall CER results for both existing and new languages under the languages-agnostic scenario. 
In this experiment, we utilize our best-performing model, where we set the LoRA rank to 512 and initiate the dual-pipeline from the intermediate layer 16.
We evaluate only 102 languages present in the FLEURS dataset, comprising 83 existing and 19 new languages.
For the language-agnostic mode, we employ the decoder selection strategy described in Section~\ref{sec:decoder_selection}.

Table~\ref{tab:all_CER} presents the results of experiments conducted using the decoder selection strategy with various values for the threshold ($\tau$) and bias score ($\beta$). 
We observed that setting $\tau = 0.5$ and$\beta = 0.15$ resulted in the best overall average CER of 29.81\%, representing a slight degradation from the group-aware result. 
As anticipated, increasing the bias score enhances the performance of the 19 new languages, albeit with a negative impact on the performance of the 83 existing languages.

Interestingly, in the language-agnostic mode, we noted that the performance on existing languages outperforms the group-aware scenario for $\beta$ set to 0 and 0.15. 
Upon analyzing the results, we found that this is attributed to a significant improvement in some low-resource languages, such as Hausa and Somali, where over 50\% absolute CER improvements are achieved.
Additionally, per-language CER for the 19 new languages is reported in Table 4.

\begin{table}[t]
    \caption{
    Average WER results for 19 new and 83 existing languages across the two scenarios: group-aware and language-agnostic, utilizing various thresholds ($\tau$) and bias scores ($\beta$) for the decoder selection strategy.}
    \label{tab:all_CER}
    \vspace{-0.2cm}
    \centering
    \small
    \renewcommand\arraystretch{0.9}
    \setlength{\tabcolsep}{2.5mm}
    \begin{tabular}{lccc}
        \toprule
                                            & \scell{19 new} & \scell{83 existing} & \scell{All}\\ 
        \midrule
        Group-aware                         & \textbf{11.18}     & 33.95     & \textbf{29.71} \\
        Language-agnostic                   &           &           & \\
        \ \ $\tau = 0.5$ \& $\beta = 0$     & 22.47     & \textbf{32.32}     & 30.48 \\
        \ \ $\tau = 0.5$ \& $\beta = 0.15$  & 16.35     & 32.90     & 29.81 \\
        \ \ $\tau = 0.5$ \& $\beta = 0.3$   & 11.89     & 35.70     & 31.27 \\
        \bottomrule
    \end{tabular}
   \vspace{-0.2cm}
\end{table} 

\begin{table}[t]
    \caption{CER results on 19 new languages from the FLEURS using zero-shot, group-aware, and language-agnostic operation modes. In the language-agnostic scenario, $\tau = 0.5$ and $\beta = 0.3$ were used in decoder selection strategy.}
    \label{tab:cer_results}
    \vspace{-0.2cm}
    \centering
    %\small
    \footnotesize
    \renewcommand\arraystretch{1.0}
    \setlength{\tabcolsep}{1mm}
    \begin{tabular}{rlS[table-format=3.2]S[table-format=2.2]S[table-format=2.2]}
        \toprule
            & {Language}         & {Zero-shot} & {Group-aware} &L{anguage-agnostic} \\
        \midrule
        1   & Asturian          & 13.32     & 7.25     & 7.25 \\
        2   & Cebuano           & 17.08     & 5.09     & 5.09 \\
        3   & Fula              & 98.36     & 17.43    & 17.44 \\
        4   & Ganda             & 48.76     & 9.76     & 9.76 \\
        5   & Igbo              & 71.63     & 13.56    & 15.00 \\
        6   & Irish             & 61.80     & 24.31    & 24.31 \\
        7   & Kabuverdianu      & 32.90     & 5.48     & 5.48 \\
        8   & Kamba             & 88.17     & 13.51    & 15.22 \\
        9   & Kyrgyz            & 34.00     & 5.91     & 5.88 \\
        10  & Luo               & 74.65     & 6.18     & 6.32 \\
        11  & Northern Sotho    & 214.07    & 8.58     & 10.09 \\
        12  & Nyanja            & 87.29     & 8.59     & 8.99 \\
        13  & Oriya             & 67.84     & 12.64    & 12.64 \\
        14  & Oromo             & 120.44    & 16.15    & 16.15 \\
        15  & Sorani Kurdish    & 77.80     & 9.34     & 9.98 \\
        16  & Umbundu           & 88.88     & 18.47    & 24.62 \\ %confused with the asturian/spanish/etc.
        17  & Wolof             & 178.77    & 15.39    & 16.77 \\ 
        18  & Xhosa             & 63.31     & 7.74     & 7.74 \\
        19  & Zulu              & 63.31     & 7.09     & 7.10\\
        \midrule
        \multicolumn{2}{l}{Average CER} & 81.62 & 11.18 & 11.89 \\
        \bottomrule
    \end{tabular}
    \vspace{-0.2cm}
\end{table}

\section{Discussion}
A substantial body of research has been conducted on fine-tuning large-scale mASR models for new languages~\cite{thomas2022efficient,chen2023exploring}. 
While these approaches are considered computationally efficient, they often result in catastrophic forgetting of existing languages, which contrasts with our objective. 
Another group of related works employs a continual learning~\cite{li2021scaling,DBLP:conf/icassp/LiPZSSHZFGP22}. 
However, these approaches rely on the availability of data from existing languages. 
Moreover, these techniques can be computationally expensive since they necessitate multiple iterations of full fine-tuning to determine optimal hyperparameters.
For instance, determining optimal data balancing weights that maintain performance on existing languages within an allowed range while achieving acceptable performance on new languages is highly challenging, especially for severely unbalanced languages.

On the other hand, our method has been demonstrated to achieve good results using as few as 0.1\% additional parameters per language, along with a relatively small number of training steps, computational resources, and training data. 
Importantly, we introduce a simple decoder selection strategy that allows us to balance performance between existing and new languages without any additional training stages. Nevertheless, we acknowledge that the main drawback of our work is the secondary decoder. Therefore, future research should focus on finding an effective way to eliminate it.

\section{Conclusion}
We proposed an effective method for integrating new languages into a pre-trained mASR system. 
The proposed method utilizes the dual-pipeline with LoRA, maintaining two pipelines for existing and new languages.
Importantly, our method does not require training data for existing languages and achieves parameter efficiency by leveraging LoRA.
Furthermore, it sustains performance on existing languages and tasks by maintaining the parameters of the pre-trained mASR unchanged.
We demonstrate the efficacy of the proposed method by extending the Whisper model to include 19 previously unseen languages.
In addition,  we compared our method to other strong baselines and implemented a basic decoder selection strategy to showcase the language-agnostic mode, achieving promising results with controllable degradation on existing languages.

\vfill
\pagebreak
\bibliographystyle{IEEEtran}
\bibliography{refs}

\begin{thebibliography}{10}
\providecommand{\url}[1]{#1}
\csname url@samestyle\endcsname
\providecommand{\newblock}{\relax}
\providecommand{\bibinfo}[2]{#2}
\providecommand{\BIBentrySTDinterwordspacing}{\spaceskip=0pt\relax}
\providecommand{\BIBentryALTinterwordstretchfactor}{4}
\providecommand{\BIBentryALTinterwordspacing}{\spaceskip=\fontdimen2\font plus
\BIBentryALTinterwordstretchfactor\fontdimen3\font minus \fontdimen4\font\relax}
\providecommand{\BIBforeignlanguage}[2]{{%
\expandafter\ifx\csname l@#1\endcsname\relax
\typeout{** WARNING: IEEEtran.bst: No hyphenation pattern has been}%
\typeout{** loaded for the language `#1'. Using the pattern for}%
\typeout{** the default language instead.}%
\else
\language=\csname l@#1\endcsname
\fi
#2}}
\providecommand{\BIBdecl}{\relax}
\BIBdecl

\bibitem{DBLP:conf/icml/RadfordKXBMS23}
A.~Radford, J.~W. Kim, T.~Xu, G.~Brockman, C.~McLeavey, and I.~Sutskever, ``Robust speech recognition via large-scale weak supervision,'' in \emph{International Conference on Machine Learning (ICML)}, vol. 202.\hskip 1em plus 0.5em minus 0.4em\relax {PMLR}, 2023, pp. 28\,492--28\,518.

\bibitem{zhang2023google}
Y.~Zhang, W.~Han, J.~Qin, Y.~Wang, A.~Bapna, Z.~Chen, N.~Chen, B.~Li, V.~Axelrod, G.~Wang \emph{et~al.}, ``Google {USM}: Scaling automatic speech recognition beyond 100 languages,'' \emph{arXiv preprint arXiv:2303.01037}, 2023.

\bibitem{pratap2023scaling}
V.~Pratap, A.~Tjandra, B.~Shi, P.~Tomasello, A.~Babu, S.~Kundu, A.~Elkahky, Z.~Ni, A.~Vyas, M.~Fazel-Zarandi \emph{et~al.}, ``Scaling speech technology to 1,000+ languages,'' \emph{arXiv preprint arXiv:2305.13516}, 2023.

\bibitem{DBLP:conf/iclr/HeZMBN22}
J.~He, C.~Zhou, X.~Ma, T.~Berg{-}Kirkpatrick, and G.~Neubig, ``Towards a unified view of parameter-efficient transfer learning,'' in \emph{International Conference on Learning Representations (ICLR)}, 2022.

\bibitem{DBLP:conf/icml/HoulsbyGJMLGAG19}
N.~Houlsby, A.~Giurgiu, S.~Jastrzebski, B.~Morrone, Q.~de~Laroussilhe, A.~Gesmundo, M.~Attariyan, and S.~Gelly, ``Parameter-efficient transfer learning for {NLP},'' in \emph{International Conference on Machine Learning (ICML)}, vol.~97.\hskip 1em plus 0.5em minus 0.4em\relax {PMLR}, 2019, pp. 2790--2799.

\bibitem{DBLP:conf/acl/MahabadiR0H20}
R.~K. Mahabadi, S.~Ruder, M.~Dehghani, and J.~Henderson, ``Parameter-efficient multi-task fine-tuning for transformers via shared hypernetworks,'' in \emph{Proceedings of the Association for Computational Linguistics and the International Joint Conference on Natural Language Processing ({ACL/IJCNLP}), (Volume 1: Long Papers)}, 2021, pp. 565--576.

\bibitem{DBLP:journals/nn/ParisiKPKW19}
G.~I. Parisi, R.~Kemker, J.~L. Part, C.~Kanan, and S.~Wermter, ``Continual lifelong learning with neural networks: {A} review,'' \emph{Neural Networks}, vol. 113, pp. 54--71, 2019.

\bibitem{DBLP:conf/icassp/LiPZSSHZFGP22}
B.~Li, R.~Pang, Y.~Zhang, T.~N. Sainath, T.~Strohman, P.~Haghani, Y.~Zhu, B.~Farris, N.~Gaur, and M.~Prasad, ``Massively multilingual {ASR:} {A} lifelong learning solution,'' in \emph{International Conference on Acoustics, Speech and Signal Processing (ICASSP)}.\hskip 1em plus 0.5em minus 0.4em\relax {IEEE}, 2022, pp. 6397--6401.

\bibitem{li2021scaling}
B.~Li, R.~Pang, T.~N. Sainath, A.~Gulati, Y.~Zhang, J.~Qin, P.~Haghani, W.~R. Huang, M.~Ma, and J.~Bai, ``Scaling end-to-end models for large-scale multilingual {ASR},'' in \emph{Automatic Speech Recognition and Understanding Workshop (ASRU)}.\hskip 1em plus 0.5em minus 0.4em\relax IEEE, 2021, pp. 1011--1018.

\bibitem{pratap20c_interspeech}
V.~Pratap, A.~Sriram, P.~Tomasello, A.~Hannun, V.~Liptchinsky, G.~Synnaeve, and R.~Collobert, ``{Massively Multilingual ASR: 50 Languages, 1 Model, 1 Billion Parameters},'' in \emph{Proc. Interspeech}, 2020, pp. 4751--4755.

\bibitem{DBLP:conf/iclr/HuSWALWWC22}
E.~J. Hu, Y.~Shen, P.~Wallis, Z.~Allen{-}Zhu, Y.~Li, S.~Wang, L.~Wang, and W.~Chen, ``{LoRA}: Low-rank adaptation of large language models,'' in \emph{International Conference on Learning Representations ({ICLR})}, 2022.

\bibitem{DBLP:conf/slt/ConneauMKZADRRB22}
A.~Conneau, M.~Ma, S.~Khanuja, Y.~Zhang, V.~Axelrod, S.~Dalmia, J.~Riesa, C.~Rivera, and A.~Bapna, ``{FLEURS:} few-shot learning evaluation of universal representations of speech,'' in \emph{Spoken Language Technology Workshop (SLT)}.\hskip 1em plus 0.5em minus 0.4em\relax {IEEE}, 2022, pp. 798--805.

\bibitem{zhang2018accelerating}
B.~Zhang, D.~Xiong, and J.~Su, ``Accelerating neural transformer via an average attention network,'' in \emph{Association for Computational Linguistics (ACL)}, 2018, pp. 1789--1798.

\bibitem{DBLP:conf/icassp/ChanJLV16}
W.~Chan, N.~Jaitly, Q.~V. Le, and O.~Vinyals, ``Listen, attend and spell: {A} neural network for large vocabulary conversational speech recognition,'' in \emph{International Conference on Acoustics, Speech and Signal Processing (ICASSP)}.\hskip 1em plus 0.5em minus 0.4em\relax {IEEE}, 2016, pp. 4960--4964.

\bibitem{ba2016layer}
J.~L. Ba, J.~R. Kiros, and G.~E. Hinton, ``Layer normalization,'' \emph{arXiv preprint arXiv:1607.06450}, 2016.

\bibitem{DBLP:conf/cvpr/HeZRS16}
K.~He, X.~Zhang, S.~Ren, and J.~Sun, ``Deep residual learning for image recognition,'' in \emph{Computer Vision and Pattern Recognition ({CVPR})}.\hskip 1em plus 0.5em minus 0.4em\relax {IEEE}, 2016, pp. 770--778.

\bibitem{khassanov_icassp24}
Y.~Khassanov, Z.~Chen, T.~Chen, T.~Y. Chong, W.~Li, L.~Lu, and Z.~Ma, ``Extending multilingual asr to new languages using supplementary encoder and decoder components,'' in \emph{International Conference on Acoustics, Speech and Signal Processing (ICASSP)}.\hskip 1em plus 0.5em minus 0.4em\relax IEEE, 2024, pp. 10\,586--10\,590.

\bibitem{DBLP:conf/nips/VaswaniSPUJGKP17}
A.~Vaswani, N.~Shazeer, N.~Parmar, J.~Uszkoreit, L.~Jones, A.~N. Gomez, L.~Kaiser, and I.~Polosukhin, ``Attention is all you need,'' in \emph{Advances in Neural Information Processing Systems ({NIPS})}, 2017, pp. 5998--6008.

\bibitem{DBLP:conf/aaai/WangCG20}
C.~Wang, K.~Cho, and J.~Gu, ``Neural machine translation with byte-level subwords,'' in \emph{The Association for the Advancement of Artificial Intelligence ({AAAI})}.\hskip 1em plus 0.5em minus 0.4em\relax {AAAI} Press, 2020, pp. 9154--9160.

\bibitem{DBLP:journals/corr/KingmaB14}
D.~P. Kingma and J.~Ba, ``Adam: {A} method for stochastic optimization,'' in \emph{International Conference on Learning Representations (ICLR)}, 2015.

\bibitem{thomas2022efficient}
B.~Thomas, S.~Kessler, and S.~Karout, ``Efficient adapter transfer of self-supervised speech models for automatic speech recognition,'' in \emph{International Conference on Acoustics, Speech and Signal Processing (ICASSP)}.\hskip 1em plus 0.5em minus 0.4em\relax IEEE, 2022, pp. 7102--7106.

\bibitem{chen2023exploring}
Z.-C. Chen, C.-L. Fu, C.-Y. Liu, S.-W.~D. Li, and H.-y. Lee, ``Exploring efficient-tuning methods in self-supervised speech models,'' in \emph{Spoken Language Technology Workshop (SLT)}.\hskip 1em plus 0.5em minus 0.4em\relax IEEE, 2023, pp. 1120--1127.

\end{thebibliography}
%\bibliography{strings,refs}
%\bibliography{mybib}

\end{document}